SPECIAL TOPIC — Optical field manipulation

# Quantum plasmon enhanced nonlinear wave mixing in graphene nanoflakes*

Hanying Deng (邓寒英)[1†], Changming Huang (黄长明)[2], Yingji He (何影记)[1], and Fangwei Ye (叶芳伟)[3‡]

[1] *School of Photoelectric Engineering, Guangdong Polytechnic Normal University, Guangzhou 510665, China*
[2]*Department of Electronic Information and Physics, Changzhi University, Changzhi, Shanxi 046011, China*
[3]*School of Physics and Astronomy, Shanghai Jiao Tong University, Shanghai 200240, China*

A distant-neighbor quantum-mechanical method is used to study the nonlinear optical wave mixing in graphene nanoflakes (GNFs), including sum- and difference-frequency generation, as well as four-wave mixing. Our analysis shows that molecular-scale GNFs support quantum plasmons in the visible spectrum region, and significant enhancement of nonlinear optical wave mixing is achieved. Specifically, the second- and third-order wave-mixing polarizabilities of GNFs are dramatically enhanced, provided that one (or more) of the input or output frequencies coincide with a quantum plasmon resonance. Moreover, by embedding a cavity into hexagonal GNFs, we show that one can break the structural inversion symmetry and enable otherwise forbidden second-order wave mixing, which is found to be enhanced by the quantum plasmon resonance too. This study reveals that the molecular-sized graphene could be used in the quantum regime for nanoscale nonlinear optical devices and ultrasensitive molecular sensors.

## 1. Introduction

Plasmons, collective free-electron oscillations in conducting media that can concentrate light into atomic length scales, have found a wide range of applications, including optical metamaterials[1-5], nanophotonic lasers and amplifiers[6], nanoantennas[7], quantum optics[8], photovoltaic devices[9-11] and biological sensing[12]. Plasmons can boost nonlinear response due to their ability to dramatically amplify the electromagnetic fields [13-16]. Noble metals have attracted much attention because of their ability to support plasmons. It is demonstrated that plasmonic properties of noble metals are related with the size, shape and surrounding environment[17,18]. However, the practical applications of metal plasmons are severely restricted by their large ohmic loss and low tunability[19,20].

---

* Project supported by the National Natural Science Foundation of China (Grant No. 11947007), the Natural Science Foundation of Guangdong Province (Grant No. 2019A1515011499) and the Department of Education of Guangdong Province (Grant No. 2019KTSCX087).
† Corresponding author. E-mail: dhy0805@alumni.sjtu.edu.cn
‡ Corresponding author. E-mail: fangweiye@sjtu.edu.cn

In contrast, graphene[21] is an outstanding plasmonic material due to their ability to support low loss[22] and tunable plasmons[23,24]. Graphene plasmons have been found in nanoribbons[25], nanodisks[26], nanorings[27], and other confined morphologies at terahertz and mid-infrared frequencies[28,29]. Graphene plasmons in near-infrared and visible regions are highly demanded in a wide range of applications too. This might be achieved by using graphene nanoflakes (GNFs), and it is demonstrated that plasmon resonance frequencies in a GNF having lateral size $D$ scale as $\omega_P \propto \sqrt{\mathrm{E_F}/D}$, with Fermi energy $\mathrm{E_F}$ taking typical experimentally value ≤ $1\mathrm{eV}$[30]. Thus, extending graphene plasmons to the visible regime can be achieved by reducing the scale of GNFs.

Thanks to the recent progress in the chemical synthesis method, GNFs with size down to molecular-scale can be manufactured[31], allowing one to investigate plasmonic phenomena at visible-light spectrum region. Further, molecular-scale GNFs offer a tantalizing route to explore plasmonic phenomena in the quantum regime. In particular, when the geometrical size of GNFs shrinks down to the molecular regime, quantum effects play an important role in plasmon resonances[32–34]. At such scale, full quantum mechanical descriptions are required[24,33–35].

The combination of plasmonics with quantum mechanics, known as quantum plasmonics[34,36,37], has become a rapidly developing research field in recent years. The origin of classical and quantum plasmons is fundamentally different. Classical plasmons occur in systems with size larger than the electron mean free path[38], which are well described by Maxwell's equations. However, in GNFs with size down to molecular scale the electronic motion is restricted by quantum confinement, and in such system the strong charge oscillations, namely, quantum plasmons, are caused by transitions between largely delocalized quantum states[34,39].

In addition to their remarkable plasmonic properties, graphene also feature strong nonlinear optical response due to their anharmonic charge-carrier dispersion relation[40–42]. Indeed, recent experiments on optical Kerr effect[43,44], second and third harmonic generation[45–47] confirm already strong nonlinear optical effects in this material. Due to their remarkably field enhancement, graphene plasmons can further enhance the intrinsically intense nonlinearity of graphene. Such field-enhanced nonlinearity is even more prominent in molecular-scale GNFs due to their even more confined plasmons. More importantly, due to the reduction in the size, molecular GNFs have an additional benefit, that is, even at lower Fermi energy one is able to reach a high frequency range, which relaxes the demanding on a high Fermi level to access the visible-frequency range.

For GNFs with a size of tens of nanometers, the optical response is dominated by the intraband transitions and we can use a classical nonlinear conductivity to describe their nonlinear optical properties[48]. However, for GNFs with the size down to molecular scale, due to the significant influence of nonlocal and finite-size effects[32,49] on the optical response, the classical electrodynamic description fails. Thus, tight-binding (TB) model has been proposed to describe the electronic states of nanostructured graphene[24,32,41,50]. However, TB model only includes the interaction of nearest-neighbor atoms, and thus it is not accurate enough, or even invalid to describe, for example, two non-tightly bond GNFs. Previous study of the nonlinear optical wave mixing in nanostructured graphene has been based on the simple TB model[41]. The influence of the quantum cavity, size and shape of molecular GNFs on their nonlinear optical wave mixing has not been studied yet.

In this work, we use a distant-neighbor quantum-mechanical (DNQM) approach to investigate the nonlinear optical wave mixing in molecular-scale GNFs, including sum- and difference-frequency generation, as well as four-wave mixing. We consider the interactions of $\pi$-orbitals of each carbon atom between core potential of all atoms. Comparison with the TB model that only including the interactions of nearest-neighbor atoms, DNQM method is more general and accurate, because it includes the electron-core couplings of all carbon atoms in the GNFs. We demonstrate that efficient wave mixing is achieved in a molecular GNF when the input or output frequencies is close to its quantum plasmon. Moreover, our calculations indicate that the presence of quantum cavity can break the inversion symmetry of hexagonal GNFs, thus enabling plasmon-enhanced second-order wave mixing.

## 2. Calculation of nonlinear optical wave mixing in GNFs

The GNFs we considered in general are dozens of carbon atoms that are hexagonally positioned in the *x*-*y* planar. For instance, seen in Fig. 1(a) is a triangular-shaped GNF, with a lattice constant $a$ = 0.142 nm and a side length of 1.2 nm. With this size scale, this GNF contains only 46 carbon atoms, enabling us to use DNQM method to compute the nonlinear wave mixing polarizabilities of such GNFs.

We initiate our calculation by computing the electronic states of GNFs. The optical response of GNFs is dominated by the electrons in the $\pi$-bond forming $p_z$ orbitals. Thus, we use a single $p_z$ orbital to represents a carbon atom[51]. This is a universally adopted treatment for GNFs modeling, however, in contrast to previous method, we include the coupling of the $p_z$ electrons and cores of all atoms in the GNFs. We then obtain the Hamiltonian operator for a single electron and the eigenstates of GNFs can be consequently calculated by solving the Schrödinger equation (see Appendix A for more details). Finally, we calculate the polarizabilities of GNFs with a perturbative method (see Appendices A and B for more details). The GNF structures considered in our work are presented in Fig.1(b).

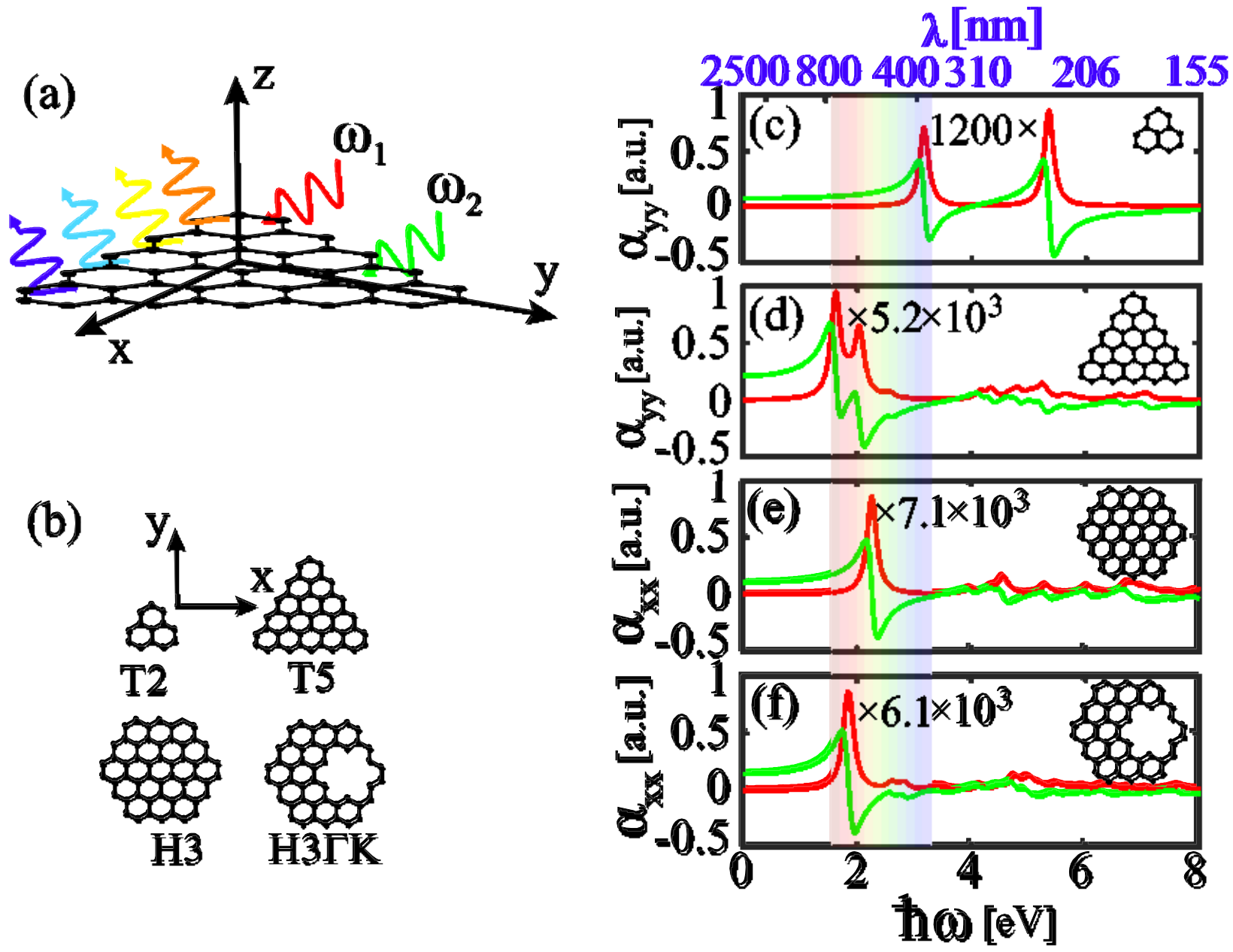

**Fig. 1 GNF structures and their linear polarizabilities.** (a) Nonlinear wave mixing in a triangular GNF, illuminated by two continuous waves with frequencies $\omega_1$ and $\omega_2$. (b) Geometrical configuration of GNFs, including Triangular (Tn) and hexagonal (Hn) structures considered in our work. Here n indicates the number of 6-atom hexagons on the side of GNFs. H3 structure containing a cavity at the $\Gamma$K direction is labeled as H3$\Gamma$K. (c)-(f) Real and imaginary parts of linear polarizability for (c) T2, (d) T5, (e) H3 and (f) H3$\Gamma$K nanoflakes. The green (red) lines indicate the real (imaginary) part of the polarizabilities. The atomic units (a. u.) with $a_0 = m_e = \hbar = e = 1$ are used in the calculation of polarizabilities of GNFs.

## 3. Results and discussion

### 3.1. Quantum plasmons of GNFs.

We first calculate the linear polarizability of molecular-scale GNFs as shown in Fig. 1(b). For triangular GNFs, we assume the incident electric field to be y-polarized and use the DNQM method to compute the linear polarizability, $\alpha_{yy}(\omega)$. The obtained imaginary and real parts of $\alpha_{yy}(\omega)$ is show in Fig. 1(c) for T2, and in Fig.1(d) for T5, respectively. The polarizability spectra clearly show that the peak of the imaginary part coincides with the zero-valued real part, which defines the existence of quantum plasmons [33, 34, 36]. A comparison between Figs. 1 (c) and 1(d) tells that the frequency of quantum plasmon profoundly increases with the decreasing of side length of triangular GNF from 1.2 nm (for T5, the first resonance peak is at 1.627 eV) to 0.49 nm (for T2, the first resonance peak is at 3.16 eV).

The dependence of linear polarizability on incident frequency, $\alpha_{xx}(\omega)$, of hexagonal GNFs with and without a quantum cavity is presented in Figs. 1(e) and 1(f), respectively. Here, the polarization direction of incident electric field is along the *x*-axis and a cavity (see inset of Fig. 1(f)) is created by removing a few carbon atoms along the $\Gamma\mathrm{K}$ symmetry axis. It is seen that the cavity can dramatically alter the optical response of GNFs. Particularly, by comparing the linear polarizability spectra presented in Figs. 1(e) and 1(f), we find that the existence of a cavity into the H3 nanoflake induces a red-shift of quantum plasmon frequencies. More importantly, all molecular-scale GNFs shown in Fig. 1(b) have quantum plasmon resonances in the visible spectral ranges, without requiring electron doping.

### 3.2. Sum- and difference-frequency generation in GNFs.

We next study the second-order nonlinear wave mixing, including sum- and difference-frequency generation, in molecular-scale GNFs. Notice that, for hexagonal GNFs, its centrosymmetric structure means no second or other even-order nonlinear processes could occur. In contrast, second-order nonlinear processes of non-centrosymmetric triangular GNFs is enabled when, for example, an incident electric field polarized along the asymmetric *y* direction is applied. Thus we focus on sum- and difference-frequency generation in triangular GNFs for a *y*-polarized applied field.

In contour plots Figs. 2(a) and 2(b), sum-frequency generation (SFG) polarizabilities $\beta_{yyy}(\omega_1+\omega_2)$ of T2 and T5 nanoflakes are shown as a function of two incident frequencies $(\omega_1,\omega_2)$. In order to see how the nonlinear wavelength mixing is connected to the plasmon resonance, in the upper panels in Fig. 2 we also present the linear polarizability $\alpha_{yy}(\omega)$ of the considered triangular GNFs. Remarkably, as shown in Figs. 2(a) and 2(b), the applied field frequencies at which the SFG is enhanced, are found to coincide exactly with the plasmon resonances shown in $\alpha_{yy}(\omega)$ (for example, the T2 structure has an enhanced SFG at $\hbar\omega_{1,2}=3.16$ or $5.35$ eV, which is also the very two frequencies at which plasmon resonance occurs). Moreover, it can be seen that the SFG enhancement also occurs at the output frequency defined by the curve $\omega_1 + \omega_2 = \omega_p$, $\omega_p$ being one of the plasmon frequencies of the GNF. Finally, a very strong enhancement is observed where the above two features intersect.

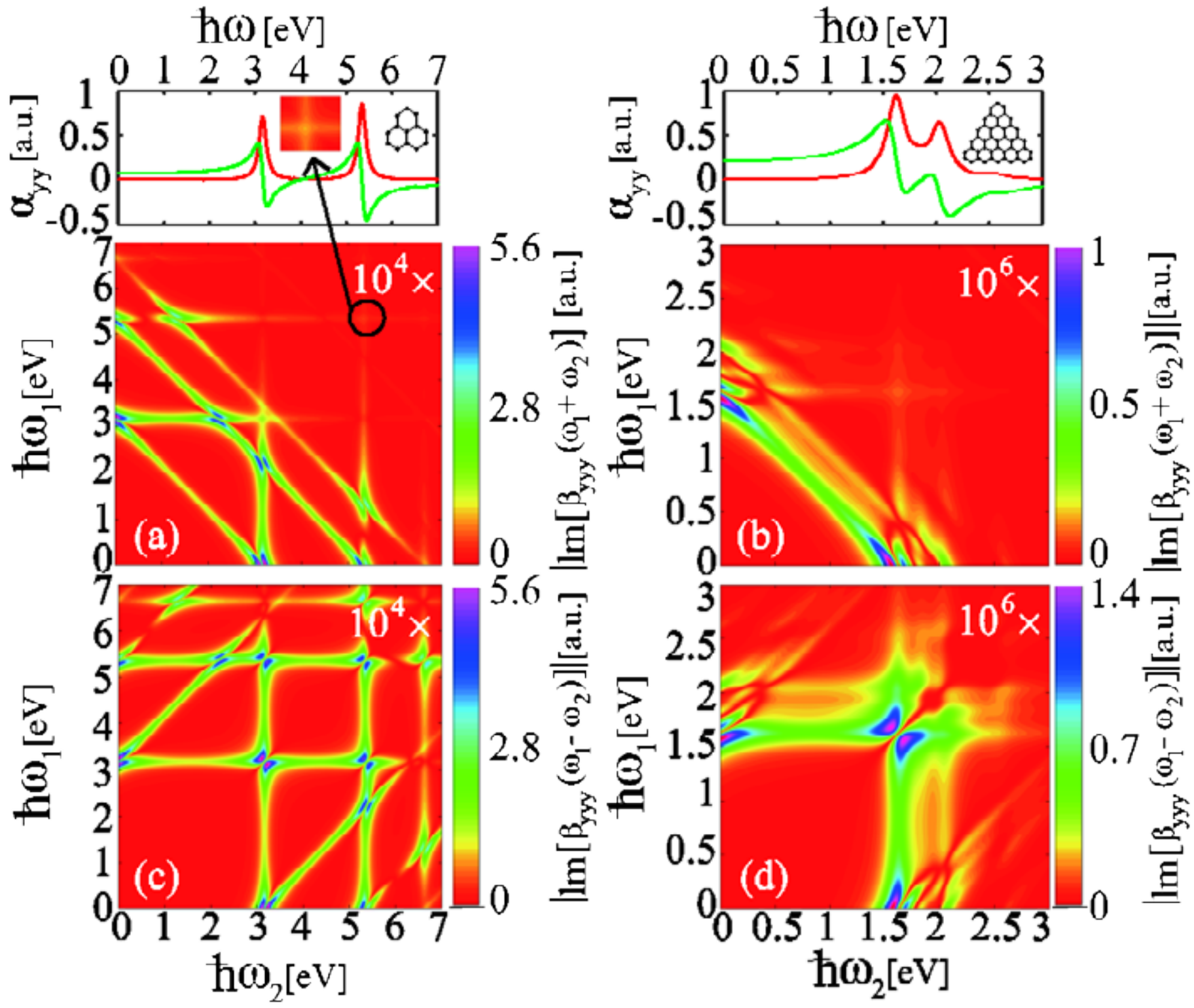

**Fig. 2. Sum- and difference-frequency generation in triangular GNFs.** (a), (b) Nonlinear polarizability $\beta_{yyy}(\omega_1 + \omega_2)$, corresponding to sum-frequency generation (SFG), for (a) T2 and (b) T5. (c), (d) Nonlinear polarizability $\beta_{yyy}(\omega_1 - \omega_2)$, corresponding to difference-frequency generation (DFG), for (c) T2 and (d) T5. The upper panels show the linear polarizability of triangular GNFs, T2 (left) and T5 (right). The inset in the top left panel shows the enlarged plot of SFG polarizability when $\omega_1$ and $\omega_2$ close to 5.35eV.

Figures 2(c) and 2(d) present the second-order polarizability, $\beta_{yyy}(\omega_1 - \omega_2)$, associated with the difference-frequency generation (DFG). Similarly, enhancement in DFG appears when one of the incident frequencies coincides with the plasmon resonance, as well as when the output frequency follows the frequencies $\omega_1 - \omega_2 = \omega_p$. The results revealed here suggest a unique route to boost various nonlinear responses in GNFs.

### 3.3. Four-wave mixing in GNFs.

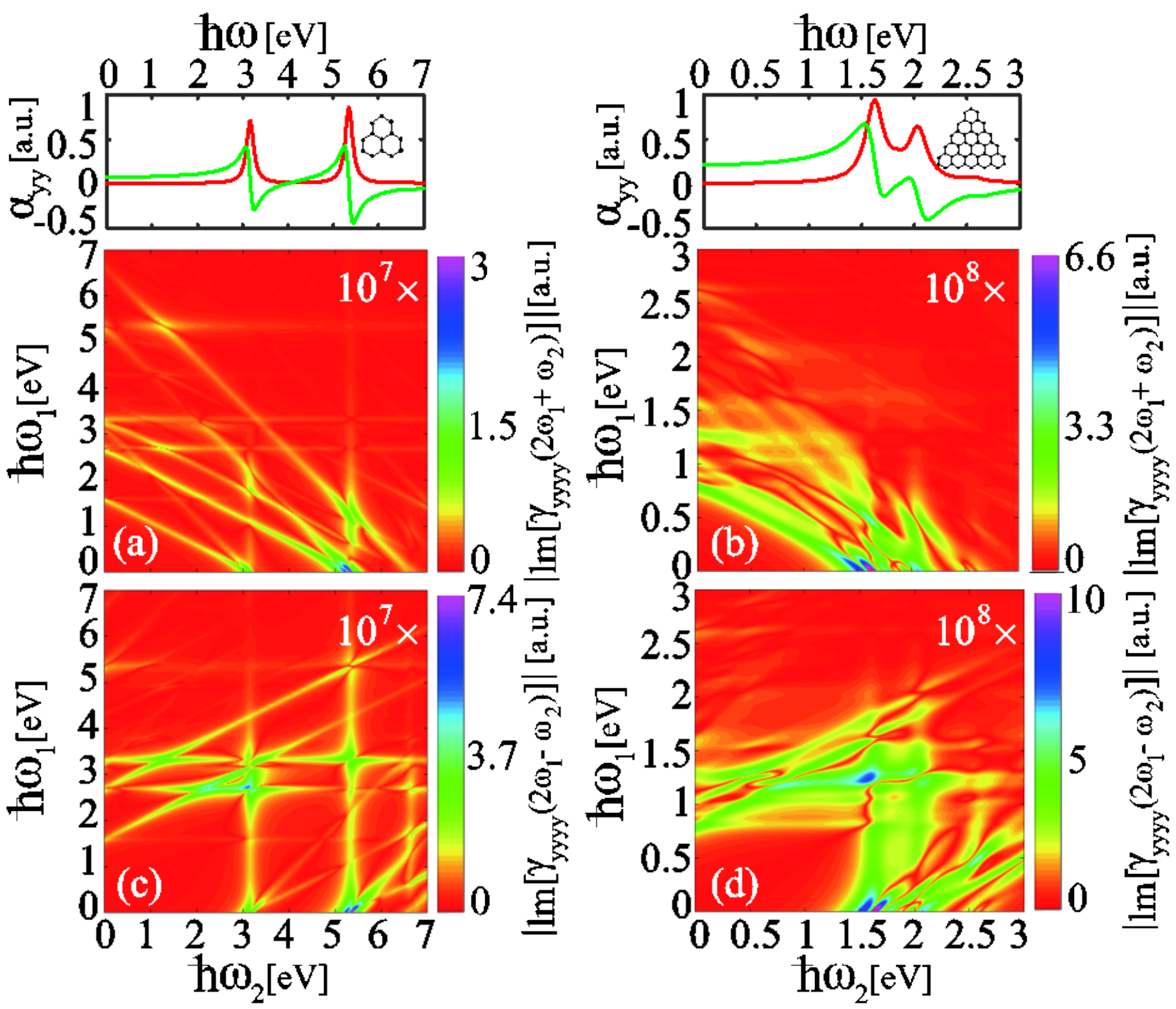

**Fig. 3 Four-wave mixing in triangular GNFs.** We show wave mixing polarizabilities (a, b) $\gamma_{yyyy}(2\omega_1+\omega_2)$ and (c, d) $\gamma_{yyyy}(2\omega_1-\omega_2)$ for the triangular GNFs considered in Fig. 2.

We now use DNQM approach to explore the four-wave mixing (FWM) processes between two applied field with frequencies $\omega_1$ and $\omega_2$ in GNFs. In Figs. 3(a) and 3(b), we present the third-order nonlinear polarizabilities corresponding to FWM with output frequency $2\omega_1+\omega_2$, $\gamma_{yyyy}(2\omega_1+\omega_2)$, for triangular GNF T2 and T5, respectively. It can be clearly seen that the enhancement of FWM at frequencies following $2\omega_1+\omega_2=\omega_p$, indicating that the output frequency resonates with a plasmon. We have also examined FWM with output frequency $2\omega_1-\omega_2$ in the triangular GNFs, the corresponding polarizabilities of T2 and T5 nanoflakes being presented in Figs. 3(c) and 3(d). Similarly, we can see that enhancement at the output frequency follows the curve $2\omega_1 - \omega_2 = \omega_p$. Thus, Figure 3 illustrates that the enhancement of FWM occurs whenever the fundamental frequencies or the incident fields are resonant with plasmons (see horizontal and vertical features in the contour plots).

In Fig. 4, we present FWM polarizabilities of hexagonal GNF H3 as a function of the two incident frequencies. The top panels show the linear polarizability of H3 nanoflake. Similar to the case of triangular GNFs, as can be seen from the $\gamma_{xxxx}(2\omega_1 + \omega_2)$ and $\gamma_{xxxx}(2\omega_1 - \omega_2)$ spectra shown in Figs. 4(a) and 4(b), efficient FWM is achieved in the hexagonal GNF when one or more of the incident or the mixed frequencies coincide with its quantum plasmon, indicating quantum plasmon-assisted enhancement of nonlinear wave mixing.

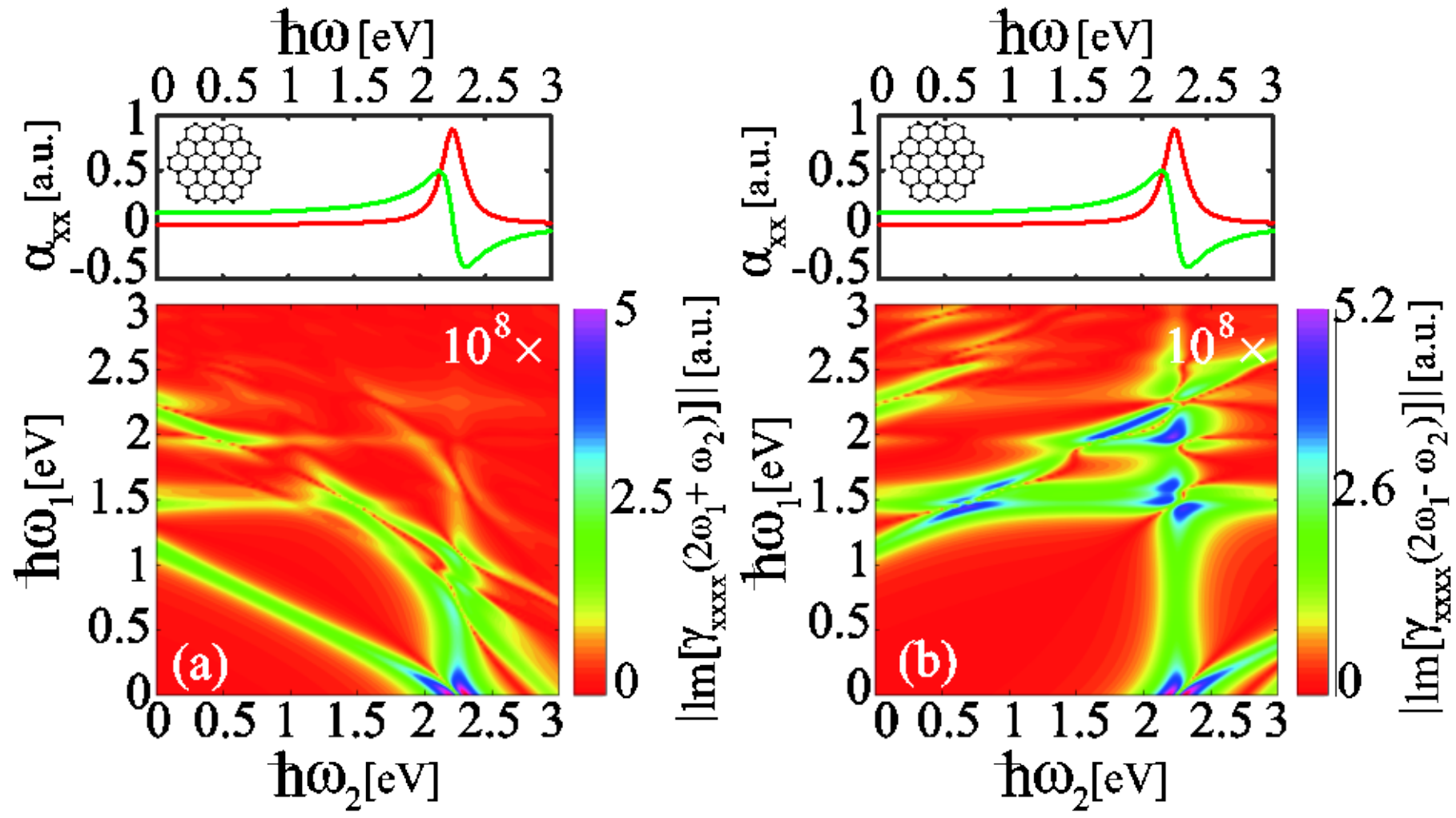

**Fig. 4 Four-wave mixing in hexagonal GNFs.** We show four-wave mixing polarizabilities (a) $\gamma_{xxxx}(2\omega_1 + \omega_2)$ and (b) $\gamma_{xxxx}(2\omega_1 - \omega_2)$ of the hexagonal GNF H3. The upper panels show the corresponding linear polarizability.

### 3.4. Nonlinear wave mixing in GNFs containing a quantum cavity.

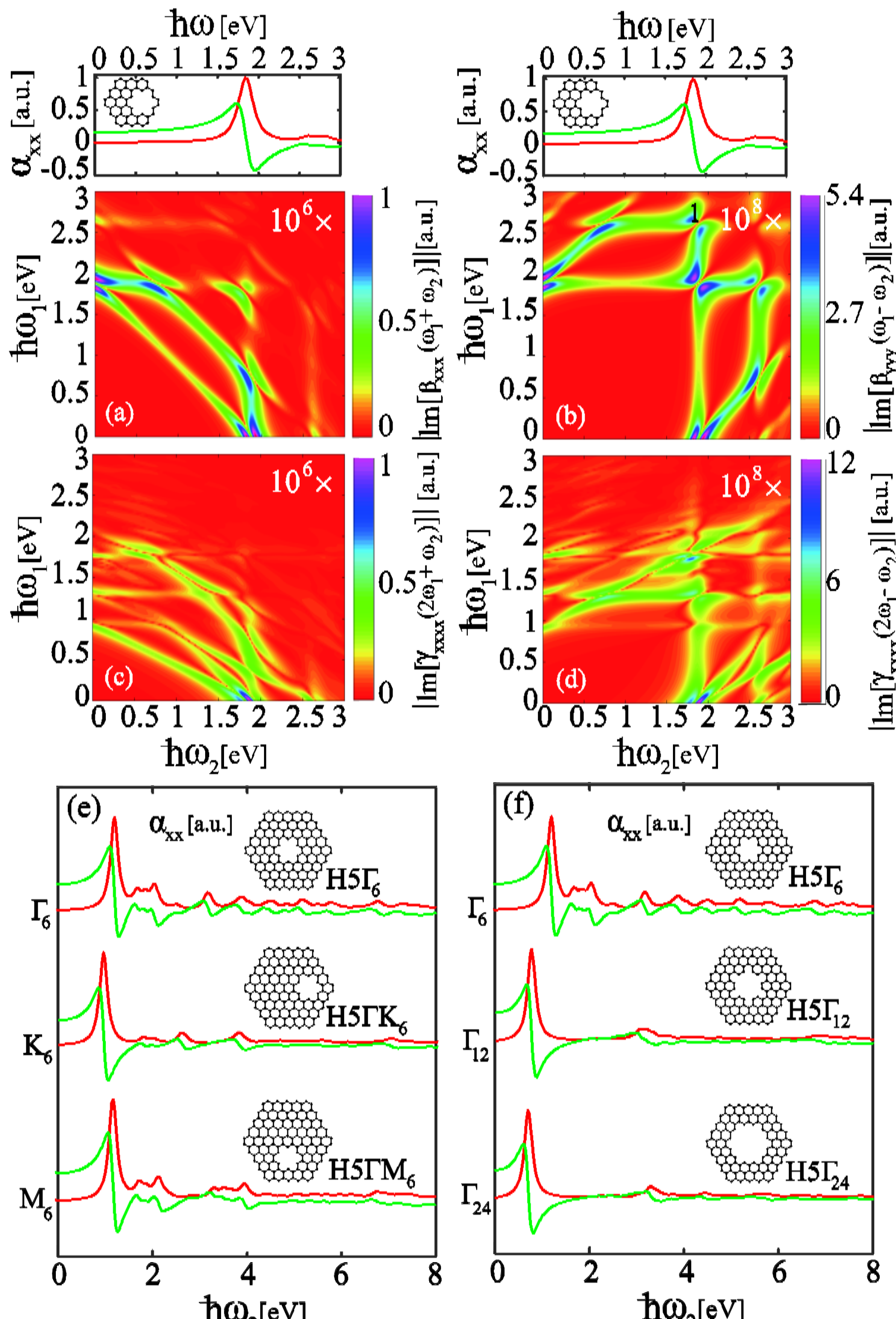

**Fig. 5 Nonlinear wave mixing in hexagonal GNFs containing a quantum cavity.** We consider the hexagonal GNF H3 with a cavity along the $\Gamma K$ direction, and show (a) the sum-frequency generation polarizability $\beta_{xxx}(\omega_1+\omega_2)$, (b) the difference-frequency generation polarizability $\beta_{xxx}(\omega_1-\omega_2)$ and the four-wave mixing polarizabilities (c) $\gamma_{xxxx}(2\omega_1+\omega_2)$ and (d) $\gamma_{xxxx}(2\omega_1-\omega_2)$. (e) The linear polarizability of H5 GNFs containing a 6-atoms cavity at the $\Gamma$ point, and along $\Gamma K$ and $\Gamma M$ directions. (f) The linear polarizability of H5 GNFs containing a cavity centered at $\Gamma$ point, introduced by removing 6 , 12 and 24 carbon atoms from the GNF, respectively, from top to bottom.

Finally, we computed the second- and third-order wave mixing in the hexagonal GNF containing a quantum cavity along the $\Gamma K$ symmetry axis, including sum- and difference-frequency generation, as well as four-wave mixing. The results of these calculations are presented in Fig. 5. As mentioned in the preceding section, the second-order nonlinear processes are forbidden in hexagonal GNFs due to the inversion symmetry of such structures. However, as shown in Figs. 5(a) and 5(b), strong sum and difference-frequency generation is enabled in the hexagonal GNF by embedding a cavity into the structure, and the quantum plasmon-assisted enhancement in $\beta_{xxx}(\omega_1+\omega_2)$ and $\beta_{xxx}(\omega_1-\omega_2)$ is clearly observed.

Figures 5(c) and 5(d) present third-order wave mixing polarizabilities $\gamma_{xxxx}(2\omega_1+\omega_2)$ and $\gamma_{xxxx}(2\omega_1-\omega_2)$, corresponding to four-wave mixing with output frequency $2\omega_1+\omega_2$ and $2\omega_1-\omega_2$, respectively. We can clearly see that the FWM polarizability spectra of H3 nanoflake with and without a cavity are markedly different, implying that the quantum cavity can efficiently alter the nonlinear polarizabilities of GNFs. Moreover, we show the effect of the position and size of the cavity on the linear polarizabilities of H5 hexagonal GNFs in Figs. 5(e) and 5(f), respectively. Figure 5(e) shows that the first plasmon frequency of H5 nanoflake with a cavity in the ΓK direction redshifts than that with a cavity at the Γ point or along ΓM directions, while Figure 5(f) shows the first resonance peak redshifts when the cavity size increases.

## 4. **Conclusion**

In conclusion, we have studied the nonlinear wave mixing in molecular-scale GNFs using the DNQM approach. Our analysis has revealed that the nonlinear response of molecular GNFs is significantly enhanced by quantum plasmon resonances, when one or more of the input or output frequencies couple to plasmons. Moreover, quantum plasmonic response and plasmon-induced enhanced nonlinearities of GNFs with size down to molecular scale can be obtained within visible spectral range. We have also demonstrated that quantum plasmon-enhanced second-order wave mixing is enabled in hexagonal GNFs by introducing a quantum cavity to break the structural inversion symmetry. The synergetic combination of the large intrinsic optical nonlinearity in GNFs and the strong local field enhancement produced by their quantum plasmons holds strong potential for nonlinear nanophotonics, quantum optics and molecular sensors.

**Appendix A: Calculation of electronic states of GNFs**

The optical behavior of GNFs is determined by the electrons in the $\pi$-bond forming $p_z$ orbitals, as shown in Fig. A1(a), and each carbon atoms has one such electron. Thus, we use the $2p_z$ Clementi orbital[51] as a basis function on each atom, which is given by

$$\psi_{2p_z} = R_{2p}(r) Y_{2p_z}(\theta,\varphi), \tag{A1}$$

where the radial part is

$$R_{2p}(r) = \frac{1}{\sqrt{6}} \frac{Zr}{na_0} Z^{\frac{3}{2}} e^{-\frac{Zr}{na_0}} \tag{A2}$$

and the angular part is

$$Y_{2p_z}(\theta,\varphi) = \sqrt{\frac{3}{4\pi}} \cos\theta = \sqrt{\frac{3}{4\pi}} \frac{z}{r}. \tag{A3}$$

Here, $Z = 3.136$ is the effective nuclear charge for the $2p_z$ orbital of a carbon atom[51], $n$ is the orbital number ($n$ = 2 in this case) and $a_0$ is the Bohr radius.

For a single electron in the GNF, the Hamiltonian operator can be written as

$$\hat{H} = -\frac{\hbar}{2m} \nabla_{\vec{r}}^2 - \sum_{\alpha=1}^{N} \frac{Z_{eff} e^2}{\vec{r} - \vec{r}_{0\alpha}} \tag{A4}$$

where the value of effective core charge, $Z_{eff} = 0.637$, has been adjusted so that the computed HOMO-LUMO gap of the simplest GNF structure (i.e. benzene) is 6 eV[50]. We can obtain the electronic states from the Schrödinger equation

$$\hat{H} \psi(\vec{r}) = E \psi(\vec{r}) \tag{A5}$$

where $\psi(\vec{r})$ and $E$ and are the eigenfunction and eigenenergy. This eigenfunction is expanded in the atom centered basis functions as

$$\psi(\vec{r}) = \sum_{q=1}^{N} c_q \psi_q(\vec{r}) = \sum_{q=1}^{N} c_q \psi_{2p_z}(\vec{r} - \vec{r}_{oq}) \tag{A6}$$

Combining Eq. (A5) and Eq. (A6) yields:

$$\sum_{q=1}^{N} \hat{H} c_q \psi_q = E \sum_{q=1}^{N} c_q \psi_q \tag{A7}$$

We then multiply each side of this equation from the left by $\psi_s^*(\vec{r})$ and integrate over all space:

$$\sum_{q=1}^{N} c_q \int \psi_s^* \hat{H} \psi_q d\vec{r} = E \sum_{q=1}^{N} c_q \int \psi_s^* \psi_q d\vec{r} \,. \tag{A8}$$

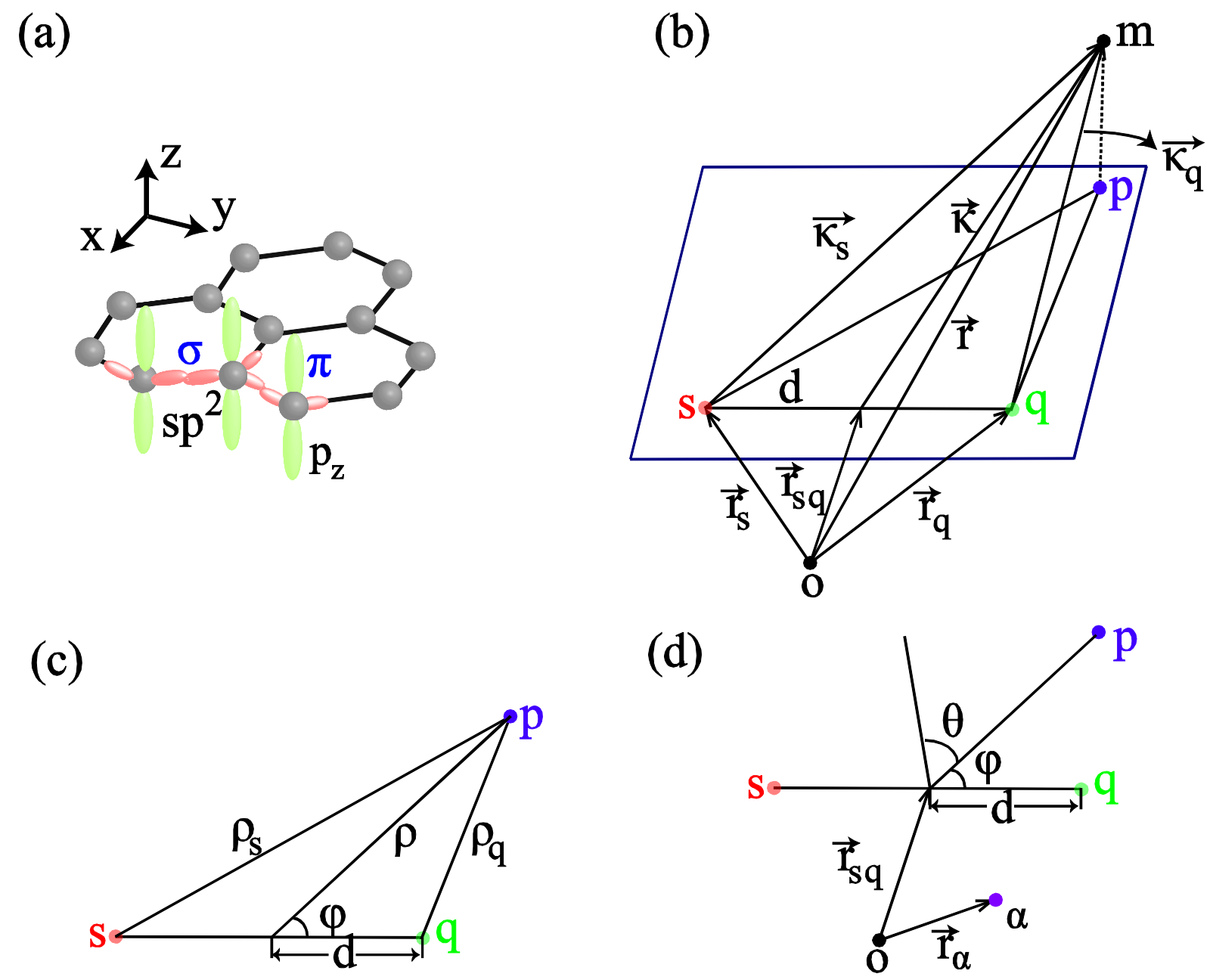

**Fig. A1** (a) Illustration of $sp^2$ hybridization in graphene. (b), (c), (d) Definitions of the vectors used in the calculations

Now we define the matrix elements

$$\hat{H}_{sq} = \int \psi_s^* \hat{H} \psi_q d\vec{r} \,, \tag{A9a}$$

$$\hat{S}_{sq} = \int \psi_s^* \psi_q d\vec{r} \,, \tag{A9b}$$

so that the Schrödinger equation is truncated to a finite generalized eigenvalue problem with eigenvectors, $\hat{c}$, and eigenvalues, $E$

$$\hat{H}\hat{c} = E\hat{S}\hat{c} \tag{A10}$$

where $\hat{c} = (c_1, c_2, ..., c_N)^T$ is an $N$-dimensional column vector.

The matrix element $\hat{H}_{sq}$ is written as the sum of two-center and three-center terms

$$\hat{H}_{sq} = E_0 \hat{S}_{sq} - \hat{I}_{sq} - \hat{I}_{sq}^{\alpha} \tag{A11}$$

where $E_0 = \dfrac{Z_{eff}\varepsilon_0}{n^2}$, and $\varepsilon_0 = -13.6\,\text{eV}$.

$$\hat{I}_{sq} = \frac{Z_{eff}e^2}{2}\int\left(\frac{1}{\left|\vec{r}-\vec{r}_s\right|} + \frac{1}{\left|\vec{r}-\vec{r}_q\right|}\right)\psi_s^*\psi_q d\vec{r}\,, \tag{A12}$$

and

$$\hat{I}_{sq}^{\alpha} = \sum_{\alpha \neq s,q} Z_{eff}e^2 \int \frac{\psi_s^*\psi_q}{\left|\vec{r}-\vec{r}_\alpha\right|} d\vec{r} \quad . \tag{A13}$$

Here $\vec{r}_s, \vec{r}_q$ and $\vec{r}_\alpha$ are the position vectors of the carbon atoms labeled by $s$, $q$, and $\alpha$, respectively.

The energy levels of a GNF containing *N* carbon atoms are abtained by solving Eq. (A10). Figures A2(a) and A2 (b) present the energy levels of the two triangular GNFs, T2 and T5. We also show the energy levels of hexagonal GNFs with and without a quantum cavity in Figs. A2(c) and A2(d), respectively.

In what follows, we describe the approach used to evaluate the multi-center integrals. We convert the spherical coordinates to a cylindrical system, as per Figs. A1(b)-A1(d), where the lengths of vectors $\vec{\kappa}_s$ and $\vec{\kappa}_q$ are expressed as

$$\left|\vec{\kappa}_s\right| = \sqrt{\rho^2 + d^2 + 2\rho d\cos(\varphi) + z^2}\,, \tag{A14a}$$

$$\left|\vec{\kappa}_q\right| = \sqrt{\rho^2 + d^2 - 2\rho d\cos(\varphi) + z^2}\,, \tag{A14b}$$

where $\rho$, $\varphi$ and $z$ are the cylindrical coordinates and $d$ is the length of vector **d**, namely it is half of the distance between the carbon atoms *s* and *q*. Therefore, the basis functions centered at *s* and *q* are written as:

$$\psi_s = Aze^{-\frac{Z}{na_0}\left|\vec{\kappa}_s\right|}\,, \tag{A15a}$$

$$\psi_q = Aze^{-\frac{Z}{na_0}\left|\vec{\kappa}_q\right|}\,, \tag{A15b}$$

where $A = \dfrac{1}{\sqrt{8\pi}}\dfrac{Z}{na_0}\left(\dfrac{Z}{a_0}\right)^{\frac{3}{2}}$ is a normalization constant. Therefore, $\hat{S}_{sq}$, $\hat{I}_{sq}^{\alpha}$, and $\hat{I}_{sq}$ are expressed as follows:

$$\hat{S}_{sq} = \int \mathrm{A}^2 z^2 e^{-\frac{Z}{na_0}\left(\left|\hat{\kappa}_s\right| + \left|\hat{\kappa}_q\right|\right)} \rho d\rho d\varphi d\theta\,, \tag{A16}$$

$$\hat{I}_{sq}^{\alpha}=\sum_{\alpha\neq s,q}\frac{A^{2}z^{2}e^{-\frac{Z}{na_0}(|\hat{\kappa}_s|+|\hat{\kappa}_q|)}}{\sqrt{\rho^{2}+\left|\vec{R}_{sq}^{\alpha}\right|^{2}-2\rho\left|\vec{R}_{sq}^{\alpha}\right|\cos\theta+z^{2}}}\rho d\rho d\varphi d\theta\,, \qquad \text{(A17)}$$

And

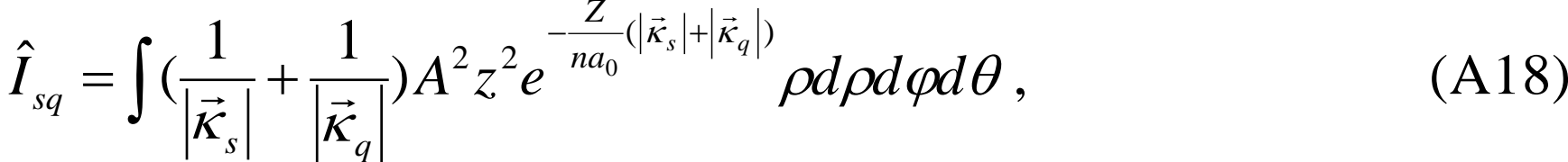

$$\hat{I}_{sq}=\int\left(\frac{1}{|\vec{\kappa}_s|}+\frac{1}{|\vec{\kappa}_q|}\right)A^{2}z^{2}e^{-\frac{Z}{na_0}(|\vec{\kappa}_s|+|\vec{\kappa}_q|)}\rho d\rho d\varphi d\theta\,, \qquad \text{(A18)}$$

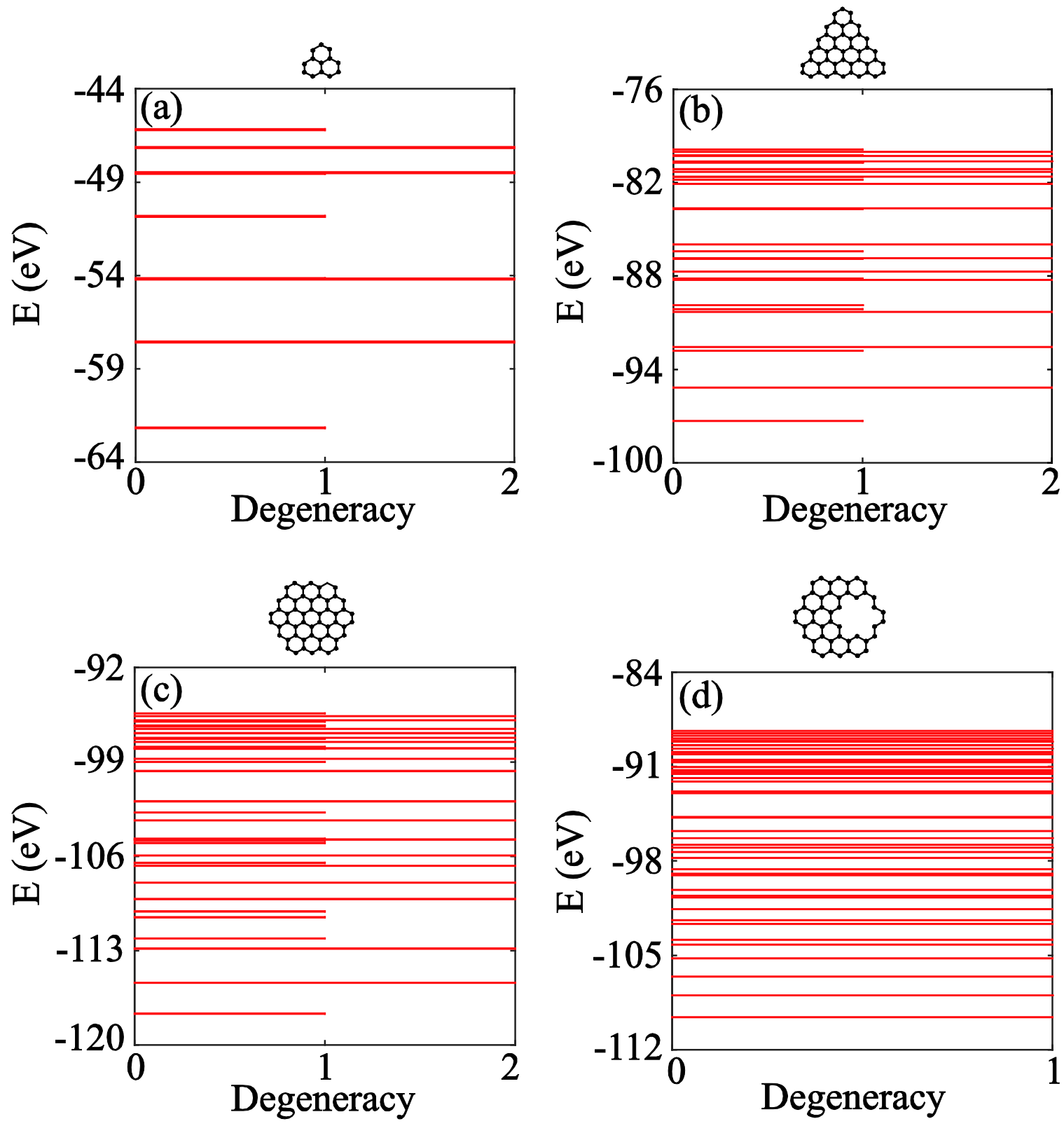

**Fig. A2** (a), (b) Energy levels of triangular GNF (a) T2 and (b) T5. (c), (d) Energy levels of hexagonal GNFs with and without a quantum cavity.

where $\left|\vec{\mathrm{R}}_{\mathrm{sq}}^{\alpha}\right|$ is the length of vector $\vec{\mathrm{R}}_{\mathrm{sq}}^{\alpha}=\vec{r}_{sq}-\vec{r}_{\alpha}$ and $\theta$ is the angle between the vector $\vec{r}$ and the plane of the graphene lattice, The integrals Eq. (A16), Eq. (A17), and Eq. (A18) have been evaluated numerically.

## Appendix B: Quantum perturbative approach to the linear and nonlinear optical response of GNFs

We decrible the optical response of GNFs using a well-known quantum perturbative approach[35]. The linear polarizability, $\alpha_{\mathrm{ij}}(\omega_k)$ ($k$=1 or 2), is given by

$$\alpha_{\mathrm{ij}}(\omega_k)=\sum_{g,m}^{transition}\left(\frac{\mu_{gm}^{i}\mu_{mg}^{j}}{\omega_{mg}-\omega_k}+\frac{\mu_{gm}^{j}\mu_{mg}^{i}}{\omega_{mg}^{*}+\omega_k}\right) \tag{A19}$$

The second-order wave-mixing polarizabilities corresponding to sum- and difference-frequency generation $\beta_{\mathrm{ijk}}(\omega_1+\omega_2)$ and $\beta_{\mathrm{ijk}}(\omega_1-\omega_2)$ are expressed as follows:

$$\begin{aligned}\beta_{\mathrm{ijk}}(\omega_1+\omega_2)=\mathrm{P}_l\sum_{g,m,n}^{transition}\Bigg[&\frac{\mu_{gn}^{i}\mu_{nm}^{j}\mu_{mg}^{k}}{(\omega_{ng}-\omega_1-\omega_2)(\omega_{mg}-\omega_1)}+\frac{\mu_{gn}^{j}\mu_{nm}^{i}\mu_{mg}^{k}}{(\omega_{ng}^{*}+\omega_2)(\omega_{mg}-\omega_1)}\\&+\frac{\mu_{gn}^{j}\mu_{nm}^{k}\mu_{mg}^{i}}{(\omega_{ng}^{*}+\omega_2)(\omega_{mg}^{*}+\omega_1+\omega_2)}\Bigg]\end{aligned} \tag{A20}$$

$$\begin{aligned}\beta_{\mathrm{ijk}}(\omega_1-\omega_2)=\mathrm{P}_l\sum_{g,m,n}^{transition}\Bigg[&\frac{\mu_{gn}^{i}\mu_{nm}^{j}\mu_{mg}^{k}}{(\omega_{ng}-\omega_1+\omega_2)(\omega_{mg}-\omega_1)}+\frac{\mu_{gn}^{j}\mu_{nm}^{i}\mu_{mg}^{k}}{(\omega_{ng}^{*}-\omega_2)(\omega_{mg}-\omega_1)}\\&+\frac{\mu_{gn}^{j}\mu_{nm}^{k}\mu_{mg}^{i}}{(\omega_{ng}^{*}-\omega_2)(\omega_{mg}^{*}+\omega_1-\omega_2)}\Bigg]\end{aligned} \tag{A21}$$

The four-wave mixing polarizabilities $\gamma_{\mathrm{kjih}}(2\omega_1+\omega_2)$ and $\gamma_{\mathrm{kjih}}(2\omega_1-\omega_2)$ are given by

$$\begin{aligned}\gamma_{\mathrm{kjih}}(2\omega_1+\omega_2)=P_l\sum_{g,m,n,,v}^{transition}\Bigg[&\frac{\mu_{gv}^{k}\mu_{vn}^{j}\mu_{nm}^{i}\mu_{mg}^{h}}{(\omega_{vg}-\omega_2-2\omega_1)(\omega_{ng}-2\omega_1)(\omega_{mg}-\omega_1)}\\&+\frac{\mu_{gv}^{j}\mu_{vn}^{k}\mu_{nm}^{i}\mu_{mg}^{h}}{(\omega_{vg}^{*}+\omega_2)(\omega_{ng}^{*}-2\omega_1)(\omega_{mg}-\omega_1)}+\frac{\mu_{gv}^{j}\mu_{vn}^{i}\mu_{nm}^{k}\mu_{mg}^{h}}{(\omega_{vg}^{*}+\omega_2)(\omega_{\mathrm{ng}}^{*}+\omega_2+\omega_1)(\omega_{mg}-\omega_1)}\\&+\frac{\mu_{gv}^{j}\mu_{vn}^{i}\mu_{nm}^{h}\mu_{mg}^{k}}{(\omega_{vg}^{*}+\omega_2)(\omega_{ng}^{*}+\omega_2+\omega_1)(\omega_{mg}^{*}+\omega_2+2\omega_1)}\Bigg]\end{aligned} \tag{A22}$$

$$\begin{aligned}\gamma_{\mathrm{kjih}}(2\omega_1-\omega_2)=P_l\sum_{g,m,n,,v}^{transition}\Bigg[&\frac{\mu_{gv}^{k}\mu_{vn}^{j}\mu_{nm}^{i}\mu_{mg}^{h}}{(\omega_{vg}+\omega_2-2\omega_1)(\omega_{ng}-2\omega_1)(\omega_{mg}-\omega_1)}\\&+\frac{\mu_{gv}^{j}\mu_{vn}^{k}\mu_{nm}^{i}\mu_{mg}^{h}}{(\omega_{vg}^{*}-\omega_2)(\omega_{ng}^{*}-2\omega_1)(\omega_{mg}-\omega_1)}+\frac{\mu_{gv}^{j}\mu_{vn}^{i}\mu_{nm}^{k}\mu_{mg}^{h}}{(\omega_{vg}^{*}-\omega_2)(\omega_{\mathrm{ng}}^{*}-\omega_2+\omega_1)(\omega_{mg}-\omega_1)}\\&+\frac{\mu_{gv}^{j}\mu_{vn}^{i}\mu_{nm}^{h}\mu_{mg}^{k}}{(\omega_{vg}^{*}-\omega_2)(\omega_{ng}^{*}-\omega_2+\omega_1)(\omega_{mg}^{*}-\omega_2+2\omega_1)}\Bigg]\end{aligned} \tag{A23}$$

Here, $\omega_1$ and $\omega_2$ are the two incident field frequencies. $\vec{\mu}=e\vec{r}$ is the dipole moment operator, $\mu_{gm}=\int\psi_g^*\vec{\mu}\psi_m d\vec{r}$ is the transition dipole moment, $\omega_{mg}=\dfrac{E_m-E_g}{\hbar}-i\eta$, $|\psi_m\rangle=\sum_l c_{ml}|\psi_l\rangle$, $E_m$ and $c_m=\sum_l c_{ml}$ are the eigenenergy and engenvector of the eigenstate *m*, respectively. Moreover, *g*, *m*, as well as *n*, *v* are labels used to distinguish between levels of transitions and $\eta$ = 0.1eV is related to the lifetime of excited states. Here, we have introduced the intrinsic permutation operator, $P_I$, defined such that the expression that follows it is to be summed over all permutations of the Cartesian indices: *i*, *j*, *k*, and *h*.

**Appendix C: Linear and nonlinear polarizability units**

In our calculation, atomic units (a.u.) are used for the linear polarizability $\alpha$, the second-order polarizability $\beta$, and the third-order polarizability $\gamma$. The units of length and energy are nm and *eV*, respectively. Following we show the transformation between a.u. and SI units.

In Eqs. (A19) - (A23), the transition dipole moment elements are in a.u., where 1 a.u. of the dipole is 1 electron charge times the Bohr radius. Specifically, we have

$$ea_0=1.602\times10^{-19}\times0.529\times10^{-10}(Cm). \tag{A24}$$

Moreover, in Eqs. (A19)-(A23) we use *eV* as the energy unit, $\omega$ must be multiplied by $\hbar$, therefore,

$$\frac{1}{\hbar\omega}=\frac{1}{1.602\times10^{-19}}(J^{-1}). \tag{A25}$$

All the coefficients and units in Eqs. (A19)-(A23) must be multiplied together. Thus, the conversion factors from the calculated value to the SI unit for $\alpha$, $\beta$, and $\gamma$ are given by

$$\begin{aligned}[\alpha]&=\frac{(ea_0)^2}{\hbar\omega}=\frac{(1.602\times10^{-19}\times0.529\times10^{-10})^2}{1.602\times10^{-19}}C^2m^2J^{-1}\\&=4.48\times10^{-40}C^2m^2J^{-1},\end{aligned} \tag{A26}$$

$$[\beta]=\frac{(ea_0)^3}{(\hbar\omega)^2}=\frac{(1.602\times10^{-19}\times0.529\times10^{-10})^3}{(1.602\times10^{-19})^2}C^3m^3J^{-2} \\ =2.37\times10^{-50}C^3m^3J^{-2}, \tag{A27}$$

and

$$[\gamma]=\frac{(ea_0)^4}{(\hbar\omega)^3}=\frac{(1.602\times10^{-19}\times0.529\times10^{-10})^4}{(1.602\times10^{-19})^3}C^4m^4J^{-3} \\ =1.25\times10^{-60}C^4m^4J^{-3}. \tag{A28}$$

## References

[1] Boltasseva A and Atwater H A 2011 *Science* **331** 290
[2] Hess O, Pendry J B, Maier, S A, Oulton R F, Hamm J M and Tsakmakidis K L 2012 *Nat. Mater.* **11** 573
[3] Zhang S, Genov D A, Wang Y, Liu M and Zhang X 2008 *Phys. Rev. Lett.* **101** 047401
[4] Chen S, Li Z, Liu W, Cheng H and Tian J 2019 *Adv. Mater.* **31** 1802458.
[5] Li Z, Liu W, Cheng H and Chen S 2020 *Sci. China-Phys. Mech. Astron.* **63**, 284202.
[6] Berini P and De Leon I D 2011 *Nat. photonics* **6** 16
[7] Giannini, V, Fernandez-Dominguez A I, Heck S C and Maier S A 2011 *Chem. Rev.* **111** 3888
[8] Akimov A V, Mukherjee A, Yu C L, Chang D E, Zibrov A S, Hemmer P R, Park H and Lukin M D 2007 *Nature* **450** 402
[9] Atwater H A and Polman A 2010 *Nat. Mater.* **9** 205
[10] Panoiu N C and Osgood R M 2007 *Opt. Lett.* **32**, 2825
[11] Clavero C 2014 *Nat. Photonics* **8** 95
[12] Anker J N, Hall W P, Lyandres O, Shah N C, Zhao J and Van Duyne R P 2008 *Nat. Mater.* **7** 442
[13] Kim S, Jin J, Kim Y J, Park I. Y, Kim Y and Kim S W 2008 *Nature* **453** 757
[14] Ko K D, Kumar A, Fung K H, Ambekar R, Liu G L, Fang N X and Toussaint K C 2011 *Nano Lett.* **11** 61
[15] Yan L, Guan M and Meng S 2018 *Nanoscale* **10** 8600
[16] Harutyunyan H, Volpe G, Quidant R and Novotny L 2012 *Phys. Rev. Lett.* **108** 217403
[17] Gramotnev D K and Bozhevolnyi S I 2010 *Nat. Photonics* **4** 83
[18] Pelton M. and Bryant G W 2013 *Introduction to metal-nanoparticle plasmonics* (Hoboken, NJ, USA: Wiley)
[19] Johnson P B and Christy R W 1972 *Phys. Rev. B* **6**, 4370
[20] Solís D M, Taboada J M, Obelleiro F, Liz-Marzan L M and De Abajo F J G 2014 *ACS Nano* **8** 7559

[21] Novoselov K S, Geim A K, Morozov S V, Jiang D, Zhang Y, Dubonos S V, Grigorieva I V and Firsov A A 2004 *Science* **306** 666
[22] Koppens F H L, Chang D E and De Abajo F J G 2011 *Nano Lett.* **11** 3370
[23] Yan H, Li Z, Li X, Zhu W, Avouris P and Xia F 2012 *Nano Lett.* **12** 3766
[24] Cox J D and De Abajo F J G 2014 *Nat. Commun.* **5** 5725
[25] Ju L, Geng B, Horng J, Girit C, Martin M, Hao Z, Bechtel H, Liang X, Zettl A and Shen Y 2011 *Nat. Nanotechnol.* **6** 630
[26] Fang Z, Wang Y, Schlather A E, Liu Z, Ajayan P M, De Abajo F J G, Nordlander P, Zhu X and Halas N J 2014 *Nano Lett.* **14** 299
[27] Gerislioglu B, Ahmadivand A and Pala N 2017 *Opt. Mater.* **73** 729
[28] Grigorenko A N, Polini M and Novoselov K S 2012 *Nat. Photonics* **6** 749
[29] De Abajo F J G 2014 *ACS Photonics* **1** 135
[30] Chen C F, Park C H, Boudouris B W, Horng J, Geng B, Girit C, Zettl A, Crommie M F, Segalman R A, Louie S G, Wang F 2011 *Nature* **471** 617
[31] Ruffieux P, Wang S, Yang B, Sánchez-Sánchez C, Liu Jia, Dienel T, Talirz L, Shinde P, Pignedoli C A, Passerone D, Dumslaff T, Feng X, Müllen K and Fasel R 2016 *Nature* **531** 489
[32] Cox J D, Silveiro I and De Abajo F J G 2015 *Acs Nano* **10** 1995
[33] Deng H, Manrique D Z, Chen X, Panoiu N C and Ye F 2018 *APL Photonics* **3** 016102
[34] Manrique D Z, You J W, Deng H, Ye F and Panoiu N C 2017 *J. Phys. Chem. C* **121** 27597
[35] Boyd R W 2008 *Nonlinear Optics.* 3rd edn. (Academic Press, Inc., USA)
[36] Tame M S, Mcenery K R, Özdemir K, Lee J, Maier S A and Kim M S 2013 *Nat. Phys.* **9** 329
[37] Bozhevolnyi S I and Mortensen N A 2016 *Nanophotonics* **6** 1185
[38] Zayats A V, Smolyaninov I I and Maradudin I I 2005 *Phys. Rep.* **408** 131
[39] Yamamoto N, Hu C, Hagiwara S and Watanabe K 2015 *Appl. Phys. Express* **8** 045103
[40] Hendry E, Hale P J, Moger J, Savchenko A K and Mikhailov S A 2010 *Phys. Rev. Lett.* **105** 097401
[41] Cox J D and De Abajo F J G 2015 *ACS Photonics* **2** 306
[42] Cox J D and De Abajo F J G 2019 *Acc. Chem. Res.* **52** 2536
[43] Zhang H, Virally S, Bao Q, Ping L K, Massar S, Godbout N and Kockaert P 2012 *Opt. Lett.* **37** 1856
[44] Tthakur S, Semnani B , Safavi-Naeini S and Majedi A H. 2019 *Sci. Rep.* **9** 10540
[45] An Y Q, Rowe J E, Dougherty D B, Lee J U and Diebold A C 2014 *Phys. Rev. B* **89** 115310
[46] Hong S Y, Dadap J I, Petrone N, Yeh P C, Hone J and Osgood Jr. R M 2013 *Phys. Rev. X* **3** 021014
[47] Kumar N, Kumar J, Gerstenkorn, Wang R, Chiu H Y, Smirl A L, Zhao H 2013 *Phys. Rev. B* **87** 121406
[48] Mikhailov A S 2007 *Eur. Lett.* **79** 417
[49] Lundeberg M B, Gao Y, Asgari R, Tan C, Van Duppen B, Autore M, Alonso-Gonzalez P, Woessner A, Watanabe K and Taniguchi T 2017 *Science* **357** 187
[50] Ezawa M 2007 *Phys. Rev. B* **76** 245415
[51] Clementi E. and Raimondi D L 1963 *J. Chem. Phys.* **38** 2686